\documentstyle[12pt]{article}

\begin{document}

\vspace*{2cm}

\noindent{\large Enhancement of low-mass dileptons in SPS heavy-ion 
collisions:}
\noindent{\large possible evidence for dropping rho meson mass in medium}
\footnote{Work supported in part by the National Science foundation under 
Grant No. PHY-9509266 and the Department of Energy under grant No.
DE-FG02-88ER40388.}

\vskip 0.5cm

\noindent C. M. Ko$^a$, G. Q. Li$^a$, G. E. Brown$^b$, and H. Sorge$^b$

\vskip 0.5cm

\noindent $^a$Cyclotron Institute and Physics Department,
Texas A\&M University, College Station, Texas 77843, USA

\vskip 0.5cm

\noindent $^b$Department of Physics, State University of New York,
Stony Brook, NY 11794, USA

\vskip 1.0cm

Dilepton production in proton- and nucleus-induced reactions at SPS
energies is studied in the relativistic transport model using initial 
conditions determined by the string dynamics from RQMD. It is found 
that both the CERES and HELIOS-3 data for dilepton spectra in 
proton-nucleus reactions can be well described by the conventional
mechanism of Dalitz decay and direct vector meson decay. However, 
to provide a quantitative explanation of the observed dilepton spectra 
in central S+Au and S+W collisions requires contributions other than 
these direct decays. Introducing a decrease of vector meson masses 
in hot dense medium, we find that these heavy-ion data can also 
be satisfactorily explained.  We also give predictions for Pb+Au 
collisions at 160 GeV/nucleon using current CERES mass resolution 
and acceptance.

\vskip 1.0cm
\noindent{\bf 1. INTRODUCTION}
\vskip 0.5cm
 
Recent observation of the enhancement of low-mass dileptons in
central S+Au and S+W collisions at SPS/CERN energies by the CERES 
[\ref{ceres95}] and the HELIOS-3 [\ref{helios95}] collaboration, 
respectively, has generated a great deal of interest in the heavy-ion 
community. Different dynamical models, such as hydrodynamical and 
transport models, have been used to investigate this phenomenon 
[\ref{li95b}-\ref{gale95}]. Although several different mechanisms 
[\ref{li95b}-\ref{koch96}] have been put forward to explain
the observed enhancement, the most successful explanation seems
to be the decrease of vector meson masses in hot dense matter
[\ref{li95b},\ref{cass95a}]. This has been worked out by us using 
the relativistic transport model based on the extended Walecka model 
[\ref{li95b}], and was supported by the calculation of Cassing 
{\it et al.} [\ref{cass95a}] using a dropping in-medium rho meson 
mass predicted from the QCD sum rules.

In this contribution, we shall review the results from our studies
for proton- and S-induced collisions and present the predictions for 
Pb+Au collisions.

\vskip 1.0cm
\noindent{\bf 2. THE RELATIVISTIC TRANSPORT MODEL AND 
IN-MEDIUM MASSES}
\vskip 0.5cm
 
In studying medium effects in heavy-ion collisions, the relativistic 
transport model [\ref{ko87}] based on the Walecka model [\ref{qhd86}]
has been quite useful. In this model, the nucleon mass is reduced
by a scalar field while its energy is shifted by a vector
field. At high densities, the nucleon mass is small and the energy
is stored in the fields. As the system expands, the nucleon mass 
returns to its free value by getting the energy back from the scalar
field energy.  The model thus provides a thermodynamically consistent 
description of the medium effects. Applications of the model to heavy-ion
collisions at SIS/GSI energies have allowed us to extract from the
experimental data useful information on hadron properties in dense 
matter [\ref{ko95}].

In heavy-ion collisions at SPS/CERN energies, many hadrons are produced
in the initial nucleon-nucleon interactions. This is usually modeled by
the fragmentation of strings, which are the chromoelectric flux-tubes 
excited from the interacting quarks [\ref{lund}]. One successful model 
for taking into account this nonequilibrium dynamics is the RQMD model 
[\ref{sorge89}]. To extend the relativistic transport model to heavy-ion 
collisions at these energies, we have used as initial conditions the 
hadron abundance and distributions obtained from the string fragmentation 
in RQMD.  Further interactions and decays of these hadrons are then 
taken into account as in usual relativistic transport model.

\vspace{11cm}

\noindent{Figure 1. Rho meson mass at finite density and temperature.
The normal nuclear matter density is denoted by 
$\rho_0=0.16~{\rm fm}^{-3}$.}
\vskip 0.8cm

To study the effects of dropping vector meson masses 
[\ref{brown91}-\ref{asakawa93}]
on the dilepton spectrum in heavy-ion collisions, we have extended
the Walecka model from the coupling of nucleons to scalar and vector
fields to the coupling of light quarks to these fields, using the 
ideas of the meson-quark coupling model [\ref{thomas94}]
and the constituent quark model. For a system of baryons (we take the 
nucleon as an example), pseudoscalar mesons ($\pi$ and $\eta$ mesons),
vector mesons (rho and omega mesons), and the axial-vector meson ($a_1$) 
at temperature $T$ and baryon density $\rho _B$, the scalar field 
$\langle\sigma\rangle$ is determined self-consistently from 
\begin{eqnarray}
m_\sigma^2\langle \sigma\rangle &=&{4g_\sigma\over (2\pi )^3}\int d{\bf k} 
{m_N^*\over E^*_N}\Big[{1\over \exp ((E^*_N-\mu _B)/T)+1}
+{1\over \exp ((E^*_N+\mu _B)/T)+1}\Big]\nonumber\\
&+&{0.45g_\sigma\over (2\pi )^3}\int d{\bf k} {m_\eta^*\over E_\eta ^*}
{1\over \exp (E_\eta ^*/T)-1}+
{6g_\sigma\over (2\pi )^3}\int d{\bf k} {m_\rho^*\over E_\rho ^*}
{1\over \exp (E_\rho ^*/T)-1}\nonumber\\
&+&{2g_\sigma\over (2\pi )^3}\int d{\bf k} {m_\omega^*\over E_\omega ^*}
{1\over \exp (E_\omega ^*/T)-1}
+{6\sqrt 2 g_\sigma\over (2\pi )^3}\int d{\bf k} {m_{a_1}^*\over E_{a_1}^*}
{1\over \exp (E_{a_1}^*/T)-1},
\end{eqnarray}
where we have used the constituent quark model relations for the nucleon
and vector meson masses [\ref{thomas94}], i.e., $m_N^*=m_N-g_\sigma\langle 
\sigma\rangle ,~m_\rho^*\approx m_\rho-(2/3)g_\sigma\langle\sigma\rangle, 
~m_\omega^*\approx m_\omega -(2/3)g_\sigma\langle\sigma\rangle$, the quark 
structure of the $\eta$ meson in free space which leads to 
$m_\eta^*\approx m_\eta -0.45g_\sigma\langle\sigma\rangle$, and 
the Weinberg sum rule relation between the rho-meson and $a_1$ 
meson masses [\ref{wein67},\ref{kapu94}], i.e, 
$m_{a_1}^*\approx m_{a_1}-(2\sqrt 2/3)g_\sigma\langle\sigma\rangle$.
In Fig. 1, we show the density and temperature dependence of the rho meson
mass using $g_\sigma=9.58$ and $m_\sigma=550$ MeV from the Walecka model. 
 
\vskip 1.0cm
\noindent{\bf 3. DILEPTON PRODUCTION}
\vskip 0.5cm
 
The main contributions to dileptons with mass below 1.2 GeV are the
Dalitz decay of $\pi^0$, $\eta$ and $\omega$, the direct leptonic decay 
of $\rho^0$, $\omega$ and $\phi$, the pion-pion annihilation which 
proceeds through the $\rho^0$ meson, and the kaon-antikaon annihilation 
that proceeds through the $\phi$ meson.  The differential widths for 
the Dalitz decay of $\pi^0$, $\eta$, and $\omega$ are related to their 
radiative decay widths via the vector dominance model, which are taken 
from Ref. [\ref{land85}].  We do not introduce the vector dominance 
model form factor for the Dalitz decay of $a_1$ meson as the 
processes $a_1\leftrightarrow\pi\rho$ and $\rho\rightarrow l^+l^-$ 
are treated explicitly as a two-step process in our model.

The decay of a vector meson into dilepton is given by the width,
\begin{equation}\label{lwidth}
\Gamma_{V\to l^+l^-}(M)= C_{l^+l^-}
\frac{m_V^4}{3M^3}(1-\frac{4m_l^2}{M^2})^{1/2}(1+\frac{2m_l^2}{M^2}).
\end{equation}
The coefficient 
$C_{l^+l^-}$ in the dielectron channel is $8.814\times 10^{-6}$,
$0.767\times 10^{-6}$, and $1.344\times 10^{-6}$ for $\rho$, $\omega$,
and $\phi$, respectively, and is determined from the measured width.
For the dimuon channel, these values are slightly larger.

When medium effects on vector meson masses are included, $m_V$ is replaced 
by $m_V^*$.  Also, the in-medium strong decay width is calculated with 
the in-medium mass. We have neglected the collisional broadening of vector
meson widths in medium [\ref{hag95}], based on the argument that their 
magnitudes are comparable to the mass resolution in CERES experiments, 
so they do not affect appreciably the final results.

In our model, dileptons are emitted continuously during the time evolution 
of the colliding system.  The way the dilepton yield is calculated can 
be illustrated by rho meson decay.  Denoting, at time $t$, the differential 
multiplicity of neutral rho mesons by $dN_{\rho ^0} (t)/dM$, then the 
differential dilepton production probability is given by [\ref{li95a}]
\begin{eqnarray}\label{sum}
{dN_{l^+l^-}\over dM} =\int _0^{t_f} {dN_{\rho^0} (t)\over dM} 
\Gamma _{\rho^0
\rightarrow l^+l^-}(M) dt + {dN_{\rho ^0} (t_f)\over dM} 
{\Gamma _{\rho^0\rightarrow l^+l^-} (M)\over\Gamma _\rho (M)},
\end{eqnarray}
where $t_f$ is the freeze-out time, which is found to be about 20 fm/c.
The first term corresponds to dilepton emission before freeze out while
the second term is from decay of rho mesons still present at freeze out.
 
\vskip 0.5cm
\noindent{\bf 3.1. Proton-nucleus collisions}

\vspace{11cm}

\noindent{Figure 2.  Dilepton invariant mass spectra from (a) p+Be 
collisions at 450 GeV, (b) p+Au collisions at 200 GeV, and (c) 
p+W collisions at 200 GeV after including the experimental acceptance cuts 
and mass resolution. Dashed curves give dilepton spectra from different 
sources.  Experimental data in (a) and (b) from the CERES collaboration 
[\ref{ceres95}] and in (c) from the HELIOS-3 collaboration [\ref{helios95}]
are shown by solid circles, with statistical errors given by bars. Brackets 
represent the square root of the quadratic sum of systematic and 
statistical errors.}
\vskip 0.8cm

The results for dilepton spectra from p+Be collisions at 450 GeV,
p+Au collisions at 450 GeV, and p+W collisions at 200 GeV are shown
in Figs. 2(a), 2(b), and 2(c), respectively, together with data from the 
CERES [\ref{ceres95}] and the HELIOS-3 collaboration [\ref{helios95}].  
It is seen that the data can be well reproduced by Dalitz decay of 
$\pi^0$, $\eta$ and $\omega$ mesons, and direct leptonic decay of
$\rho^0$, $\omega$ and $\phi$ mesons. These results are thus similar 
to that found in Ref. [\ref{cass95a}] using the Hadron-String Dynamics
and constructed by the CERES collaboration from known and expected 
sources of dileptons [\ref{ceres95}].
 
\vskip 0.5cm
\noindent{\bf 3.2. Nucleus-nucleus collisions}

\vspace{11cm}

\noindent{Figure 3. Dilepton invariant mass spectra in (a) S+Au and 
(b) S+W collisions at 200 GeV using free (dashed curves) and in-medium 
(solid curve) meson masses. Experimental data in (a) from the CERES 
collaboration [\ref{ceres95}] and in (b) from the HELIOS-3 collaboration 
[\ref{helios95}] are shown by solid circles.}
\vskip 0.8cm

With free meson masses, the calculated dilepton spectra, normalized by the
average charged-particle multiplicity, are shown in Fig. 3(a) by the 
dashed curve together with the CERES data. Although pion-pion annihilation
is important for dileptons with invariant mass from 0.3 to 0.65 GeV, 
it still does not give enough number of dileptons in this mass region.
Furthermore, for masses around $m_{\rho,\omega}$ there are 
more dileptons predicted by the theoretical calculations than shown in 
the experimental data. These are very similar to our earlier results based 
on a thermally equilibrated fire cylinder model [\ref{li95b}] as well
as that of Cassing {\it et al.} [\ref{cass95a}] based on the 
Hadron-String Dynamics model and Srivastava {\it et al.} [\ref{gale95}]
based on the hydrodynamical model.
 
The results obtained with in-medium meson masses are shown in Fig. 3(a) 
by the solid curve.  Compared with the results obtained with free meson 
masses, there is about a factor of 2-3 enhancement of the dilepton yield 
in the mass region from 0.2 to 0.6 GeV, which thus leads to a good 
agreement between the theoretical results with the CERES data. This 
is again very similar to that found in Ref. [\ref{li95b}] using the
fire-cylinder model.
 
The same model has been used to calculate the dimuon spectra from central 
S+W collisions by the HELIOS-3 collaboration. The results obtained
with free meson masses are shown in Fig. 3(b) by the dashed curve,
and are below the HELIOS-3 data in the mass region from 0.35 to 0.6 GeV, 
and slightly above the data around $m_{\rho ,\omega}$ as in the CERES 
case. However, the discrepancy between the theory and the data is somewhat 
smaller in this case due to the smaller charged-particle multiplicity at 
a larger rapidity than in the CERES experiment.
 
Our results obtained with in-medium meson masses are shown in Fig. 3(b)
by the solid curve, and are in good agreement with the data.  The 
importance of in-medium meson masses in explaining the HELIOS-3 data 
has also been found by Cassing {\it et al.} [\ref{cass95a}].
 
\vskip 0.5cm
\noindent{\bf 3.3. Predictions for Pb+Au collisions}

\vspace{11cm}

\noindent{Figure 4. Dilepton spectra from central Pb+Au collisions
with free (dashed curve) and in-medium (solid curve) meson masses.
The CERES mass resolution and acceptance cuts for S+Au collisions
are included.}
\vskip 0.8cm
 
Dilepton production in Pb+Au collisions at 160 GeV/nucleon is currently
being measured by the CERES collaboration.  We have calculated the
dilepton spectrum for this reaction using present CERES mass 
resolution and acceptance cuts. The theoretical predictions 
for the central collisions are shown in Fig. 4 for the two scenarios 
of free (dashed curve) and in-medium meson masses (solid curve).  
The normalization factor $dN_{ch}/d\eta$ here is the average charge 
particle pseudo-rapidity density in the pseudo-rapidity range of 2 
to 3, and is about 440 in this collision. With free meson masses, there 
is a strong peak around $m_{\rho ,\omega}$, which is dominated by 
$\rho^0$ meson decay as a result of an enhanced contribution from 
pion-pion annihilation in Pb+Au collisions than in S+Au and 
proton-nucleus collisions. With in-medium meson masses, the $\rho$ 
meson peak shifts to a lower mass, and the peak around $m_{\rho ,\omega}$ 
becomes a shoulder arising mainly from $\omega$ meson decay. At the same 
time we see an enhancement of low-mass dileptons in the region of 
0.25-0.6 GeV as in S+Au collisions. 

Since the dilepton yield from both Dalitz decay and $\omega$ leptonic 
decay increases roughly linearly with the charged-particle multiplicity, 
whereas the contribution from pion-pion annihilation increases more 
than linearly, the dilepton spectra per charged particle in the invariant
mass region of $0.3<M<0.6$ GeV, where pion-pion annihilation dominates,
are somewhat larger in Pb+Au collisions than in S+Au collisions.  
However, the medium effects are similar in the two collisions. Both 
dilepton spectra show a low mass peak at about 400 MeV, and this is 
due to the similar initial rho meson mass in the two cases. The initial 
nuclear density in Pb+Au collision is about 4$\rho_0$ and is higher than 
that (about 2.5 $\rho_0$) in S+Au collisions, but the difference in the rho 
meson mass in the two cases is small as shown in Fig. 1. Furthermore, 
the yield of low mass dileptons is about the same as the dilepton
yield at the rho peak in the spectra obtained with the free rho meson 
mass. Although there are more rhos when the mass is reduced, their 
contribution to low mass dileptons is suppressed due to both a smaller 
leptonic decay width (see Eq. (\ref{lwidth})) and a short duration of
high density matter in the collision (see Eq. (\ref{sum})).

\vskip 1.0cm
\noindent{\bf 4. SUMMARY AND OUTLOOK}
\vskip 0.5cm
 
In summary, we have studied dilepton production from both proton-nucleus 
and nucleus-nucleus collisions using the relativistic transport
model with initial conditions determined by string fragmentation
from the initial stage of the RQMD model.  It is found that the dilepton 
spectra in proton-nucleus reactions measured by the CERES and the HELIOS-3 
collaboration can be well understood in terms of conventional mechanisms 
of Dalitz decay and direct vector meson decay. For dilepton spectra in 
central S+Au and S+W collisions, these conventional mechanisms, however, 
fail to explain the data, especially in the low-mass region from about 
0.25 to about 0.6 GeV in CERES experiments, and from 0.35 to 0.65 GeV 
in HELIOS-3 experiments. Including the contribution from pion-pion 
annihilation, which is important in the mass region from 
$2m_\pi$ to $m_{\rho ,\omega}$, removes some of the discrepancy.  
But the theoretical prediction is still substantially below the 
data in the low mass region and somewhat above the data around 
$m_{\rho,\omega}$. The theoretical results are brought into good 
agreement with the data when reduced in-medium vector meson masses 
are taken into account.  The results of the present study based on 
initial conditions from the RQMD model are thus very similar to our 
earlier results assuming that initially there is a thermally 
equilibrated fire-cylinder.
 
We have also presented predictions for the dilepton spectrum from central 
Pb+Au collisions. Since the medium effects in our model already saturate 
at a few times of normal nuclear density, the results from this reaction
turn out to be very similar to that from S+Au collisions. However, 
medium effects are expected to become stronger in heavy-ion collisions 
at lower incident energies when the system is initially in a soft mixed 
phase and thus expands slowly, allowing thus a longer time for the
emission of low mass dileptons [\ref{shuryak}]. 

\vskip 1.0cm
\noindent{\bf REFERENCES}
\vskip 0.5cm

\begin{list}{\arabic{enumi}.\hfill}{\setlength{\topsep}{0pt}
\setlength{\partopsep}{0pt} \setlength{\itemsep}{0pt}
\setlength{\parsep}{0pt} \setlength{\leftmargin}{\labelwidth}
\setlength{\rightmargin}{0pt} \setlength{\listparindent}{0pt}
\setlength{\itemindent}{0pt} \setlength{\labelsep}{0pt}
\usecounter{enumi}}

\item\label{ceres95} G. Agakichiev {\it et al.}, Phys. Rev. Lett. 75
(1995) 1272; J. P. Wurm for the CERES Collaboration, Nucl. Phys. 
A590, 103c (1995); I. Tserruya, Nucl. Phys. A590
(1995) 127c. 
\item\label{helios95} M. Masera for the HELIOS-3 Collaboration, Nucl. 
Phys. A590 (1995) 93c. 
\item\label{li95b} G. Q. Li, C. M. Ko, and G. E. Brown, Phys. Rev. Lett.
75 (1995) 4007; Nucl. Phys. A, in press. 
\item\label{cass95a} W. Cassing, W. Ehehalt, and C. M. Ko, Phys. Lett. 
B363 (1995) 35; W. Cassing, W. Ehehalt, and I. Kralik, {\it ibid.} 
B377 (1996) 5. 
\item\label{gale95} D. K. Srivastava, B. Sinha, and C. Gale, Phys. Rev. 
C53 (1996) R567. 
\item\label{huang95} Z. Huang and X. N. Wang, Phys. Rev. D53
(1996) 5034.  
\item\label{kapu95} J. Kapusta, D. Kharzeev, and L. McLerran, Phys. Rev. 
D53 (1996) 5028. 
\item\label{wam95} R. Rapp, G. Chanfray, and J. Wambach, Phys. Rev. Lett.
76 (1996) 368. 
\item\label{koch96} V. Koch and C. Song, LBL preprint, LBL-38619,
nucl-th/9606028. 
\item\label{ko87} C. M. Ko, Q. Li, and R. Wang, Phys. Rev. Lett. 59
(1987) 1084; C. M. Ko and Q. Li, Phys. Rev. C37 (1988) 2270; 
Q. Li, J. Q. Wu, and C. M. Ko, Phys. Rev. C39 (1989) 849;
C. M. Ko, Nucl. Phys. A495 (1989) 321c.  
\item\label{qhd86} B. D. Serot and J. D. Walecka, Adv. Nucl. Phys.
16 (1986) 1. 
\item\label{ko95} C. M. Ko and G. Q. Li, Nucl. Phys. A583 (1995)
591c. 
\item\label{lund} B. Anderson, G. Gustafson, and B. Nilsson-Almqvist, Nucl. 
Phys. B281 (1987) 289; B. Nilsson-Almqvist and E. Stenlund, 
Comp. Phys. Comm. 43 (1987) 387.
\item\label{sorge89} H. Sorge, H. St\"ocker, and W. Greiner, Ann. Phys. 
192 (1989) 266; H. Sorge,  Phys. Rev C52 (1995) 3291;
Phys. Lett. B373 (1996) 16.  
\item\label{brown91} G. E. Brown and M. Rho, Phys. Rev. Lett. 66 
(1991) 2720. 
\item\label{hatsuda92} T. Hatsuda and S. H. Lee, Phys. Rev. C46
(1992) R34. 
\item\label{asakawa93} M. Asakawa and C. M. Ko, Phys. Rev. C48
(1993) R526; Nucl. Phys. A560 (1993) 399. 
\item\label{thomas94} K. Saito and A. W. Thomas, Phys. Rev. C51
(1995) 2757. 
\item\label{wein67} S. Weinberg, Phys. Rev. Lett. 18 (1967) 507.
\item\label{kapu94} J. I. Kapusta and E. V. Shuryak, Phys. Rev. D49
(1994) 4694. 
\item\label{land85} L. G. Landberg, Phys. Rep. 128 (1985) 301.
\item\label{hag95} K. Haglin, Nucl. Phys. A584 (1995) 719. 
\item\label{li95a} G. Q. Li and C. M. Ko, Nucl. Phys. A582 (1995) 731.
\item\label{shuryak} E. Shuryak and L. Xiong, Phys. Lett. B333 (1994) 316.
 
\end{list}
 
\end{document}